\documentclass[%
 reprint,
 amsmath,amssymb,
 aps,
]{revtex4-2}

\usepackage[caption=false]{subfig}

\usepackage[colorlinks=true]{hyperref}

\usepackage{physics}
\usepackage{multirow, graphicx,amssymb,url,mathrsfs,amsmath}
\usepackage{wrapfig,boxedminipage,epsfig}
\usepackage{amsxtra,amstext,latexsym,dsfont,amsfonts}
\usepackage{color}
\usepackage[dvipsnames]{xcolor}
\usepackage{float}
\usepackage{slashed}
\usepackage{calligra}
\DeclareFontShape{T1}{calligra}{m}{n}{<->s*[2.2]callig15}{}
\DeclareMathAlphabet{\mathcalligra}{T1}{calligra}{m}{n}
\newcommand{\be}{\begin{equation}}
\newcommand{\ee}{\end{equation}}
\newcommand{\bea}{\begin{eqnarray}}
\newcommand{\eea}{\end{eqnarray}}

\begin{document}

\title{Quantum-gravitational noise correlation in nearby detectors}

\author{Maulik Parikh,}
\email{maulik.parikh@asu.edu}
\affiliation{Department of Physics, Arizona State University, Tempe, Arizona 85287, USA}
\affiliation{Beyond: Center for Fundamental Concepts in Science,
Arizona State University, Tempe, Arizona 85287, USA}

\author{Francesco Setti}
\email{fsetti@asu.edu}
\affiliation{Department of Physics, Arizona State University, Tempe, Arizona 85287, USA}

%
%\affiliation[a]{Department of Physics, Arizona State University, AZ 85287, United States}
%\affiliation[b]{Beyond: Center for Fundamental Concepts in Science,
%Arizona State University, Tempe, Arizona 85287, USA}

% e-mail addresses:
%\email{maulik.parikh@asu.edu}
%\email{fsetti@asu.edu}

\vskip 0.6in

\begin{abstract}
    \noindent We consider quantum gravity fluctuations in a pair of nearby gravitational wave detectors. Quantum fluctuations of long-wavelength modes of the gravitational field induce coherent fluctuations in the detectors, leading to correlated noise. We determine the variance and covariance in the lengths of the arms of the detectors, and thereby obtain the graviton noise correlation. We find that the correlation depends on the angle between the detector arms as well as their separation distance. Using our result, we propose an experimental setup to detect this noise. The suggested interferometer configuration can be used to distinguish quantum-gravitational noise from both correlated and uncorrelated sources of background noise. 
    %We conclude by explaining how graviton noise affects the arms of a real gravitational wave interferometer. 
\end{abstract}

%\begin{document}
\maketitle

\section{Introduction}

It is widely expected on theoretical grounds that gravity must ultimately obey the laws of quantum mechanics {\cite{Feynman:1963ax, Weinberg:1964kqu, Deser:1969wk, Boulware:1974sr}}. By contrast there is, as yet, no compelling {\em experimental} evidence for the quantization of gravity. Smoking-gun proofs of the quantization of gravity are remarkably challenging to find \cite{Carney:2023nzz}. Indeed, it is challenging enough to find any distinctive observational signature of quantum gravity, smoking-gun or otherwise. In particular, it has been argued by Dyson that single-graviton detection is virtually impossible \cite{Dyson:2013hbl,Rothman:2006fp, Boughn:2006st}. However, it is not necessary to detect individual gravitons to detect observational signatures of quantum gravity: just as Brownian noise can be observed even when collisions with individual molecules cannot, it can be that graviton noise is observable even when individual gravitons are not. In a series of papers, Parikh, Wilczek, and Zahariade (PWZ) calculated the effect of a quantized spacetime metric on a pair of free-falling particles \cite{Parikh:2020fhy, Parikh:2020kfh, Parikh:2020nrd}; they found that quantum-gravitational fluctuations induce stochastic fluctuations (i.e., noise) in the separation of the particles. The particle separation, which in classical general relativity would be governed by the geodesic deviation equation, now obeys a Langevin-like stochastic equation, a kind of quantum geodesic deviation equation. These results, which were extended by Cho and Hu \cite{Cho:2021gvg} and obtained by a different approach by Kanno $\textit{et al.}$ \cite{Kanno:2020usf, Kanno:2021gpt}, are exciting because they indicate the presence of quantum-gravitational fluctuations (“the noise of gravitons”) in the separation of the mirrors of a gravitational wave interferometer. The statistical properties of the noise depend on the quantum state of the gravitational field and can be greatly enhanced for certain classes of states, notably squeezed states. These results have been corroborated by several other authors \cite{Guerreiro:2021qgk, Haba:2020jqs, Haba:2021esn, Haba:2022rpu, Nandi:2021xtr, Sen:2023ksj, Moreira:2023hpj, Abrahao:2023lle}. In addition, PWZ's work has inspired much theoretical effort to better understand the behavior of the noise due to gravitons. For example, researchers have calculated the quantum corrections to the trajectory of a free-falling particle \cite{Chawla:2021lop}, and the quantum corrections in the Raychaudhuri equation \cite{Bak:2022oyn, Cho:2023dmh, Bak:2023wwo}. Furthermore, PWZ's formalism has been used to study the quantization of cylindrical gravitational waves \cite{He:2022tvk}, and the interest in squeezed states has led to a study on whether Virgo-LIGO data could be compatible with a gravitational field in such a quantum state \cite{Hertzberg:2021rbl}.

These calculations are the beginning of attempts to extract a precise observational signature of the quantization of gravity that experimentalists could search for. In order for the noise to be observable, however, two requirements need to be met. First, the amplitude of the noise must be large enough to be observable. As mentioned, this requires the gravitational field to be in an especially noisy state, such as a squeezed state; an open theoretical problem is determining whether there are realistic astrophysical scenarios that might give rise to such states. Second, the noise needs to be distinguishable from the numerous other sources of noise that gravitational wave interferometers are subject to. 

Although graviton noise has a characteristic spectrum (depending on the state), it can nevertheless be challenging to distinguish it from the many other types of noise present in gravitational wave detectors, such as thermal noise, seismic noise, electronic noise, and photon shot noise, among others \cite{LIGOScientific:2019hgc}. To address this problem PWZ suggested that
the quantum noise due to gravitons might be correlated in different detectors. This would help in the isolation of graviton noise because other types of noise are generically uncorrelated between different detectors. Additionally, obtaining an equation for the quantum-gravitational noise correlation might be useful in the field of gravito-optics, where the correlation of gravitational waves in different detectors has been studied for classical periodic and nonperiodic signals \cite{Jones:2019kil}. Intuitively, quantum fluctuations in long-wavelength modes of the gravitational field have the same coupling to nearby detectors and therefore lead to correlated stochastics. In this paper, we analyze the correlations in graviton noise, deriving the noise covariance, the standard deviation, and the correlation. We predict the existence of correlated quantum-gravitational noise between nearby detector arms, with the correlation dependent on the angle between the arms and their separation distance. Based on the angular dependence of the graviton noise correlation, we suggest a configuration of gravitational wave interferometers to distinguish quantum-gravitational noise from environmental noise.

\section{Derivation of the Correlation}
\subsection{Obtaining the equations of motion}

Consider a system with four masses $m_0$, $m_1$, $m_2$, and $m_3$. 
Let $m_0$ and $m_1$ be the mirrors of the first detector, and let $m_2$ and $m_3$ be the mirrors of the second detector. Let the metric be $g_{\mu \nu}=\eta_{\mu \nu} +\kappa h_{\mu \nu}$ where $\kappa^2 = 16 \pi G$ and $h_{\mu \nu}$ is the metric perturbation due to gravitons. 

We set up Fermi normal coordinates at the origin, so that $m_0$ is stationary. In addition, let $\xi^i$ be the vector from $m_0$ to $m_1$, $\eta^i$ be the vector from $m_0$ to $m_2$, and $\zeta^i$ be the vector from $m_0$ to $m_3$. Since we are in Fermi normal coordinates, the components of these vectors are proper lengths. In addition, $\xi = \sqrt{\delta_{ij}\xi^i \xi^j}=\xi^3$, is the arm length of the first detector, while we define the arm length of the second detector to be $\rho = \sqrt{\delta_{ij}\rho^i \rho^j}$, where $\rho^i = \zeta^i - \eta^i$.

\begin{figure}
    \centering
    \includegraphics[width=0.45\textwidth]{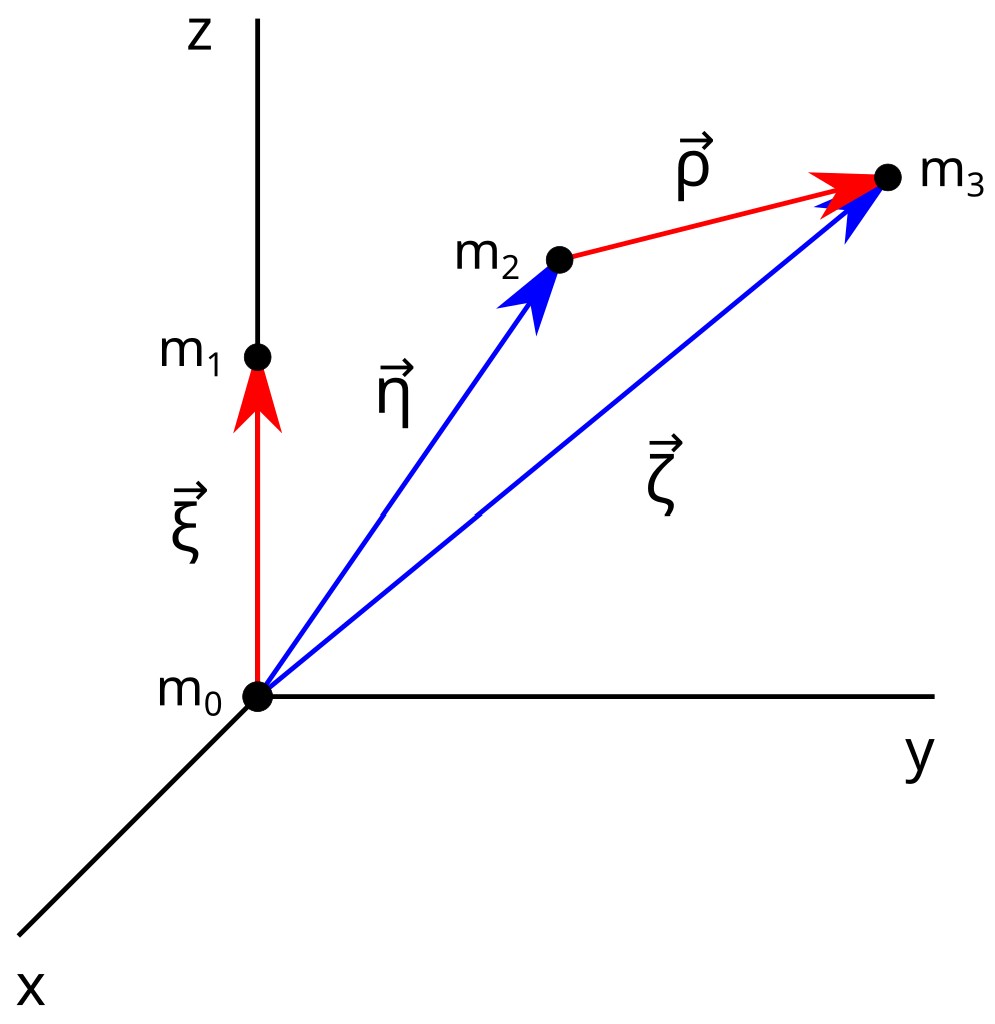}
    \caption{In the above image, $m_0$ and $m_1$ are the mirrors of the first detector, while $m_2$ and $m_3$ are the mirrors of the second detector. $\Vec{\xi}$ and $\Vec{\rho}$, shown in red, are the arms of the two detectors while $\Vec{\eta}$ and $\Vec{\zeta}$, shown in blue, are the positions of the mirrors of the second detector. Since we are working in Fermi normal coordinates, all lengths are proper lengths.}
    \label{masses and axes}
\end{figure}

Working in linearized gravity, it is straightforward to show that the classical action for the system is 

\begin{align}
\begin{split} 
\label{action}
    S_{\text{total}} =& \, S_{\text{grav}} + S_{m_0}+ S_{m_1} + S_{m_2}+ S_{m_3} \\=& \, -\frac{1}{4}  \int d^4x \,   \partial_{\alpha} h_{\mu \nu}(x) \partial^{\alpha}h^{\mu \nu}(x) \\ &  +\, \int dt  \frac{\delta_{ij}}{2} [m_1 \dot{\xi}^i \dot{\xi}^j + m_2 \dot{\eta}^i \dot{\eta}^j + m_3 \dot{\zeta}^i \dot{\zeta}^j]  \\& + \frac{\kappa}{4} \ddot{h}_{ij}(x)[m_1 \xi^i \xi^j + m_2 \eta^i \eta^j + m_3 \zeta^i \zeta^j] ,
\end{split}
\end{align}
where the metric perturbation can be decomposed into modes using the following equation
\begin{equation}
\label{modes}
    h_{\mu\nu}(x) = \int dk^3 \sum_{s= +, \cross} \epsilon_{\mu\nu}^{(s)}(\Vec{k}) \, h^{(s)}(\Vec{k},t) \, e^{i\Vec{k} \cdot \Vec{x}}
\end{equation}
Notice that in Eq. (\ref{action}) we implicitly applied a dipole approximation in order to factor out the second time derivative of the metric perturbation in the last line. This dipole approximation is needed because we are only interested in the modes that are correlated in the two detectors. Because of this approximation, the integral in Eq. (\ref{modes}) is only over the modes that have a wavelength much larger than the dimensions of the system. 

Now, suppose that the masses are initially in a quantum state $|A\rangle$ and that the gravitational field is initially in a state $|\Psi\rangle$. Then, the probability that the system is in final state $|B\rangle$ is given by the following path integral
\begin{equation}
\begin{split}
    P_{\Psi}&(A \to B) \sim  \int \mathcal{D}\xi \, \mathcal{D}\xi'\,\mathcal{D}\eta\,\mathcal{D}\eta'\mathcal{D}\zeta\,\mathcal{D}\zeta' \\ &\Bigr\{ e^{\frac{i}{\hbar} \int dt \frac{1}{2} [m_1 (\dot{\xi}^2 - \dot{\xi}'^{2})+ m_2(\dot{\eta}^2 - \dot{\eta}'^{2})+ m_3(\dot{\zeta}^2 - \dot{\zeta}'^{2})]} \\&F_{\Psi}[\xi,\xi',\eta,\eta',\zeta,\zeta'] \Bigl\}
\end{split}
\end{equation}
where $F_{\Psi}$ is the Feynman-Vernon influence functional \cite{Feynman:1963fq}. Generalizing Cho and Hu's calculations \cite{Cho:2021gvg} to the case of 4 stationary masses, we find that the influence functional can be written in terms of the noise kernel 
\begin{equation}
\label{N}
\begin{split}
    K^{\Psi}_{ijkl}(t, t') =& \frac{1}{2} \frac{d^2}{dt^2} \frac{d^2}{dt'^2} \int d^3k \, d^3k' \int d^3x \, d^3x' \\ &e^{-i \Vec{k}\cdot \Vec{x}} e^{-i \Vec{k}'\cdot \Vec{x}'} \sum_s \epsilon_{ij}^{(s)}(\Vec{k}) \epsilon_{kl}^{(s)}(\Vec{k}') G_\Psi^{(1)}(x,x')
\end{split} 
\end{equation}
with 
\begin{equation}
\label{g1}
    G_\Psi^{(1)}(x,x')=\langle \Psi|\{h(x),h(x')\}|\Psi \rangle
\end{equation} 
Next, it is possible to rewrite the part of the influence functional that involves the noise kernel using a trick due to Feynman and Vernon. As discussed in \cite{Parikh:2020fhy, Cho:2021gvg}, this introduces in the action a stochastic tensor  $\mathcal{N}_{ij}$ that has the following properties
\begin{align}
\label{Gaussian}
    &P\left[ \mathcal{N} \right]=\, \mathcal{C} e^{-\frac{1}{2}\int \mathcal{N}_{ij}[(K^\Psi)^{-1}]^{ijkl} \mathcal{N}_{kl}}\\
\label{average}
    &\expval{\mathcal{N}_{ij}(t)}_s =\int \mathcal{D}\mathcal{N} \,P\left[ \mathcal{N} \right]\mathcal{N}_{ij}(t) = \, 0 \\
\label{var}
\begin{split}
    &\expval{\mathcal{N}_{ij}(t) \mathcal{N}_{kl}(t')}_s =\int \mathcal{D}\mathcal{N} \,P\left[ \mathcal{N} \right]\mathcal{N}_{ij}(t) \mathcal{N}_{kl}(t') \\ & \quad \quad \quad \quad \quad \quad \quad = K^\Psi_{ijkl}(t,t'),
\end{split}
\end{align}
where (\ref{Gaussian}) is a Gaussian probability density, and the subscript $s$ in (\ref{average}) and (\ref{var}) denotes a statistical average. Using the new form of the action that contains the stochastic tensor, we can obtain the equations of motion by evaluating the saddle point of the path integral, which is obtained by taking a derivative of the above action with respect to $\xi$, $\eta$, or $\zeta$, and setting it equal to zero. By taking another derivative with respect to time, and dropping the radiation reaction terms, we obtain
\begin{equation}
\begin{split}
    &\ddot{\xi}^i(t)-2 \alpha \delta^{ip}\mathcal{N}_{pj}(t) \xi^j(t)=0\\& \ddot{\eta}^i(t)-2 \alpha \delta^{ip}\mathcal{N}_{pj}(t) \eta^j(t)=0\\&\ddot{\zeta}^i(t)-2 \alpha \delta^{ip}\mathcal{N}_{pj}(t) \zeta^j(t)=0
\end{split}
\end{equation}
where $\alpha= \kappa / 2 \sqrt{2} (2\pi)^3$. These are Langevin-like equations in which the arm length of the detectors is affected by the stochastic tensor. Notice that the equations of motion are uncoupled. This means that the motion of one mass does not influence the motion of the other mass. This is what we expect since, by ignoring the radiation reaction terms, we made the approximation that the masses emit no gravitational waves when they move, and so they have no way of interacting with one another. Also, this implies that any correlation in the motion of the masses is indeed due to the background gravitons.

\subsection{Solving the equations of motion to obtain the 2-point functions}
Following \cite{Bak:2022oyn}, we will solve the equations of motion perturbatively. Let us begin by solving the equation of motion for $m_1$. The equations of motion for the other masses are solved in an analogous way. 

Let $\xi^i(t)=\xi_0^i(t)+\xi_1^i(t)+\xi_2^i(t)+ \dots$, where the superscript indicates the spatial vector component and the subscripts indicate the order, which corresponds to the number of stochastic tensor factors. Then, the equation for $\xi$ becomes
\begin{align}
\ddot{\xi}^i_0(t)&=0\,,\label{eq_eom1}\\
\ddot{\xi}^i_1(t)&=2  \alpha \, \delta^{ij}\mathcal{N}_{jk}(t)\, \xi^k_0(t)\,,\label{eq_eom2}\\
\ddot{\xi}^i_2(t)&= 2 \alpha \, \delta^{ij}\mathcal{N}_{jk}(t) \, \xi^k_1(t)\,,\label{eq_eom3}
\end{align}
Integrating with respect to time, we find
\begin{align}
\label{xi_1}
    \xi^i_1(t) &= 2  \alpha \, \delta^{ij}\int^t_0 dt' \int^{t'}_0 dt'' \, \mathcal{N}_{jk}(t'') \, \xi^k_0 \\
\label{xi_2}
\begin{split}
    \xi^i_2(t) &= \left(2  \alpha \right)^2\delta^{ij} \delta^{kl}\int^t_0 dt' \int^{t'}_0 dt'' \\ & \quad \quad \int^{t''}_0 dt'''\int^{t'''}_0 dt'''' \, \mathcal{N}_{jk}(t'') \mathcal{N}_{lm}(t'''') \, \xi^m_0,
\end{split}
\end{align}
where we assumed that the unperturbed arm length $\xi^i_0$ is a constant. The equations for $\eta^i_1(t)$ and $\eta^i_2(t)$, and $\zeta^i_1(t)$ and $\zeta^i_2(t)$ are the same as (\ref{xi_1}) and (\ref{xi_2}), but with $\eta^i_0$ and $\zeta^i_0$ replacing $\xi^i_0$. 

We can now calculate the 2-point functions. It turns out that all we need to obtain the noise correlation are the 2-point functions of $\xi_1^i(t)$, $\eta_1^i(t)$, and $\zeta_1^i(t)$. In the next section, we will show why this is the case. For now, we will just calculate the various 2-point functions that we will use later. 

Since we will not use terms like $\xi^i_2(t)$ and other higher order expansion terms, we will relabel $\xi^i_1(t)$, $\eta^i_1(t)$, and $\zeta^i_1(t)$ as $\delta \xi^i(t)$, $\delta \eta^i(t)$, and $\delta \zeta^i(t)$, respectively, so that 
\begin{equation}
\begin{split}
\xi^i(t)&=\xi_0^i(t)+\delta \xi^i(t)+\mathcal{O}(\mathcal{N}^2)+\dots\,,\\
\eta^i(t)&=\eta_0^i(t)+\delta \eta^i(t)+\mathcal{O}(\mathcal{N}^2)+\dots\,,\\
\zeta^i(t)&=\zeta_0^i(t)+\delta \zeta^i(t)+\mathcal{O}(\mathcal{N}^2)+\dots\,.
\end{split} \label{expansion_FN}
\end{equation}

Let us start by finding $\expval{\delta \xi^i(t) \delta \eta^j(t)}_s$. Using Eq. (\ref{xi_1}) and (\ref{var}), we find
\begin{equation}
\label{2-point function with integrals}
\begin{split}
    &\expval{\delta \xi^i(t)  \delta \eta^j(t')}_s = (2  \alpha)^2 \, \delta^{ik} \delta^{jl} \xi_0^m \eta_0^n \\ &\quad\int^t_0 dt' \int^{t'}_0 dt'' \int^{t^{(3)}}_0 dt^{(4)} \int^{t^{(4)}}_0 dt^{(5)} K_{kmln}^\Psi(t,t'),
\end{split}
\end{equation}
where the noise kernel in the Minkowski vacuum is given by 
\begin{equation}
\label{k}
\begin{split}
     K_{ijkl}^0(t,t') =& - \frac{32 \pi^4 }{15} [2 \delta_{ij} \delta_{kl} -3 (\delta_{ik} \delta_{jl} +\delta_{il}\delta_{jk})] \\ &\int_0^\Lambda d\omega \, \omega^5 \cos{[\omega(t-t')]},
\end{split}
\end{equation}
with $\Lambda$ being the cutoff frequency. We impose a cutoff frequency because if not regularized, this integral would be divergent. Additionally, since we are only interested in the correlated modes, the cutoff is naturally $\Lambda \sim L^{-1}$, where $L$ is the largest separation between any two masses in our apparatus. For example, in Fig. \ref{masses and axes}, $L=|\Vec{\zeta}|$. We also point out that the noise kernel for different quantum states that are rotationally invariant have the same factors of Kronecker delta's present in Eq. (\ref{k}). This means that although the variance of the quantum noise is different, its angular dependency is the same. For instance, this is true for thermal states and squeezed states \cite{Cho:2021gvg,Kanno:2020usf}. Additionally, the noise kernel for a squeezed state with squeeze parameter $r$ would also be enhanced by a factor of $e^r$. Evaluating the integrals and taking the limit $t' \rightarrow t$, we find
\begin{equation}
\label{2-point function}
\begin{split}
    \expval{\delta \xi^i(t) \delta \eta^j(t)}_s &= A(\Lambda,t) \delta^{ik} \delta^{jl} \xi_0^m \eta_0^n \\&[-2 \delta_{km} \delta_{ln} +3 (\delta_{kl} \delta_{mn} +\delta_{kn}\delta_{ml})],
\end{split}
\end{equation}
where we defined the unitless time-dependent amplitude  
\begin{equation}
\label{amplitude}
    A(\Lambda,t) \simeq \frac{\kappa^2}{240 \pi^2} \Lambda^4 t^2
\end{equation}
More information about this calculation can be found in \cite{Cho:2021gvg} and \cite{Bak:2022oyn}. Since there are different ways of regularizing the integral in Eq. (\ref{k}), we only keep the first term in $A(\Lambda,t)$. More information about this can be found in our very recent paper \cite{Bak:2023wwo}.

Using Eq. $(\ref{2-point function})$ we can now find the 2-point functions for all components. Since $\xi^i(t)$ is along the $\textit{z}$-axis, we know that its only nonzero component is $\xi^3(t)$, which is also equal to its magnitude. In addition, we only need to consider $\delta \xi^3(t)$, since we cannot measure $\delta \xi^1(t)$ and $\delta \xi^2(t)$ due to the fact that these fluctuations are perpendicular to the path of the laser in the detector. Then, the only 2-point functions of interest involving $\delta \xi^i (t)$ and $\delta \eta^i (t)$ are 
\begin{equation}
\begin{split}
    \expval{\delta \xi^3(t) \delta \eta^1(t)}_s  &=-2 \, \xi_0^3 \eta_0^1\, A(\Lambda,t)\\
    \expval{\delta \xi^3(t) \delta \eta^2(t)}_s  &=-2 \, \xi_0^3 \eta_0^2 \, A(\Lambda,t)\\
    \expval{\delta \xi^3(t) \delta \eta^3(t)}_s  &= +4 \, \xi_0^3 \eta_0^3 \, A(\Lambda,t).
\end{split} 
\end{equation}
Notice that the 2-point functions for perpendicular components are proportional to $-2$, while the 2-point functions for parallel components are proportional to $+4$. In addition, all 2-point functions of $\delta \xi^i(t)$ and $\delta \eta^j(t)$ are directly proportional to corresponding factors of $\xi^i_0$ and $\eta_0^j$. Finally, all 2-point functions are proportional to the amplitude $A(\Lambda,t)$. The same is true for all other 2-point functions involving $\delta \xi^3(t)$ and $\delta \zeta^i(t)$. Specifically, 
\begin{equation}
\begin{split}
\label{rules}
    \expval{\delta \xi^3(t) \delta \zeta^1(t)}_s  &=-2 \, \xi_0^3 \zeta_0^1 \, A(\Lambda,t)\\
    \expval{\delta \xi^3(t) \delta \zeta^2(t)}_s  &=-2 \, \xi_0^3 \zeta_0^2 \, A(\Lambda,t)\\
    \expval{\delta \xi^3(t) \delta \zeta^3(t)}_s  &= +4 \, \xi_0^3 \zeta_0^3 \, A(\Lambda,t).
\end{split} 
\end{equation}

\subsection{Obtaining the noise correlation}

Our goal in this section is to obtain the graviton noise correlation in the two detectors, which is
\begin{equation}
\label{corr def}
    \text{corr}(\delta \xi, \delta \rho) = \frac{\text{cov}(\delta \xi, \delta \rho)}{\sqrt{\text{Var}(\delta \xi)} \, \sqrt{\text{Var}(\delta \rho)}}
\end{equation}
Notice that in this equation $\delta \xi$ and $\delta \rho$ are not vectors, they are magnitudes. This is because we are interested in the fluctuations in the detectors' arms, not in the fluctuations in their Cartesian components. 

Let us start by determining $\text{cov}(\delta \xi, \delta \rho)$. To do this, let us momentarily go back to our original notation where $\xi(t)=\xi_0(t)+\xi_1(t)+\xi_2(t)+ \dots$ and $\rho(t)=\rho_0(t)+\rho_1(t)+\rho_2(t)+ \dots$. By definition, we have 
\begin{align}
\begin{split}
\label{calc}
    \text{cov}( \xi(t),  \rho(t)) =& \langle[\xi(t)-\langle\xi(t)\rangle_s][\rho(t)-\langle\rho(t)\rangle_s]\rangle_s \\
    =&\langle[\xi_0(t)+\xi_1(t)+\xi_2(t)+ \dots\\&-\langle\xi_0(t)+\xi_1(t)+\xi_2(t)+ \dots\rangle_s]\\
    &\, [\rho_0(t)+\rho_1(t)+\rho_2(t)+ \dots \\&-\langle\rho_0(t)+\rho_1(t)+\rho_2(t)+ \dots\rangle_s]\rangle_s \\
    =& \langle \xi_1(t) \rho_1(t) \rangle_s + \mathcal{O}(\mathcal{N}^3) + \mathcal{O}(\mathcal{N}^4) + \dots \\
    =& \langle \xi_1(t) \rho_1(t) \rangle_s = \text{cov}( \delta \xi(t), \delta \rho(t)),
\end{split}
\end{align}
% we actually used the fact that $\expval{\xi_1(t)}_s=\expval{\rho(t)_1}_s=0$
where we used the fact that $\expval{\delta \xi(t)}_s=\expval{\delta \rho(t)}_s=0$. This means that, to leading order in the stochastic tensor $\mathcal{N}$, the covariance of the detectors arm length is equal to the covariance of the noise. The same is also true for the correlation. In addition, the above equation also demonstrated why we can neglect $\xi_2(t)$, $\rho_2(t)$, and all higher order terms.  

Next, using the fact that $\Vec{\xi}(t)$ is along the z-axis, we find 
\begin{align}
\begin{split}
\label{covariance calc}
    &\text{cov}( \delta \xi(t), \delta \rho(t))= \langle \delta \xi(t) \delta \rho(t) \rangle_s \\&= \langle \bigl[\delta \Vec{\xi}(t)\cdot \Vec{\xi}_0/\xi_0\bigr] \bigl[\delta \Vec{\rho}(t)\cdot \Vec{\rho}_0/\rho_0\bigr] \rangle_s \\
    &= \bigl(\langle \delta \xi^3(t) \delta \rho^1(t) \rangle_s \, \rho_0^1 + \langle \delta \xi^3(t) \delta \rho^2(t) \rangle_s \, \rho_0^2 \\& \quad + \langle \delta \xi^3(t) \delta \rho^3(t) \rangle_s \, \rho_0^3\bigr)/\rho_0
\end{split}
\end{align}
The 2-point functions in the above equation are evaluated exactly like the ones in the previous section. For example,
\begin{align}
\begin{split}
    \expval{\delta \xi^3(t) \delta \rho^1(t)}_s =&  \expval{\delta \xi^3(t) [\delta \zeta^1(t)-\delta \eta^1(t)]}_s \\
    =& \expval{\delta \xi^3(t) \delta \zeta^1(t)}_s - \expval{\delta \xi^3(t) \delta \eta^1(t)}_s \\
    =& -2 \, \xi^3_0 (\zeta^1_0 - \eta^1_0) \, A(\Lambda,t) \\
    =& -2 \, \xi^3_0 \rho^1_0 \, A(\Lambda,t)
\end{split}
\end{align}
With similar calculations, we determine
\begin{equation}
\begin{split}
    \expval{\delta \xi^3(t) \delta \rho^2(t)}_s  &=-2 \, \xi_0^3 \rho_0^2 \, A(\Lambda,t)\\
    \expval{\delta \xi^3(t) \delta \rho^3(t)}_s  &= +4 \, \xi_0^3 \rho_0^3 \, A(\Lambda,t)
\end{split} 
\end{equation}
Substituting these into Eq. (\ref{covariance calc}), we determine
\begin{equation}
\label{cov}
\begin{split}
    \text{cov}&( \delta \xi(t), \delta \rho(t)) \\&= \Bigl[-2 \left(\rho^1_0\right)^2 - 2 \left(\rho^2_0\right)^2 +4 \left(\rho^3_0\right)^2\Bigr]A(\Lambda,t) \, \xi_0/\rho_0
\end{split}
\end{equation}

Next, we need to find $\text{Var}(\delta \xi(t))$ and $\text{Var}(\delta \rho(t))$. With a calculation extremely similar to the one of Eq. (\ref{calc}), we determine 
\begin{equation}
\label{variance}
\begin{split}
    \text{Var}( \xi(t)) &\approx \text{Var}(\delta \xi(t)) = \expval{\delta \xi(t) \delta \xi(t)}_s \\
    \text{Var}( \rho(t)) &\approx \text{Var}(\delta \rho(t)) = \expval{\delta \rho(t) \delta \rho(t)}_s 
\end{split} 
\end{equation}
Since $\Vec{\xi}(t)$ is along the $\textit{z}$-axis,
\begin{equation}
\label{var xi}
    \text{Var}(\delta \xi(t)) = \expval{\delta \xi^3(t) \delta \xi^3(t)}_s = +4 \left(\xi_0\right)^2 A(\Lambda_\xi,t),
\end{equation}
where $\Lambda_\xi \sim \xi^{-1}$ is cutoff frequency of the first detector. Similarly, it is straightforward to show that
\begin{equation}
\label{var rho}
    \text{Var}(\delta \rho(t)) = \expval{\delta \rho(t) \delta \rho(t)}_s = +4 \left(\rho_0\right)^2 A(\Lambda_\rho,t),
\end{equation}
where $\Lambda_\rho \sim \rho_0^{-1}$ is cutoff frequency of the second detector. The details of this calculation are shown in the appendix. 

Substituting Eqs. (\ref{cov}), (\ref{var xi}), and (\ref{var rho}) into Eq. (\ref{corr def}), we determine 
\begin{equation}
\begin{split}
    \text{corr}&(\delta \xi(t), \delta \rho(t)) \\&= \frac{A(\Lambda,t)}{\sqrt{A(\Lambda_\xi,t)A(\Lambda_\rho,t)}}\frac{\left[- \left(\rho^1_0\right)^2 - \left(\rho^2_0\right)^2 + 2 \left(\rho^3_0\right)^2\right]}{2  \left(\rho_0\right)^2}
\end{split}
\end{equation}
Finally, after converting the above equation from Cartesian coordinates to spherical coordinates and writing the amplitudes $A$ in terms of the arm lengths of the detectors and the size of the system, we obtain 
\begin{equation}
\label{main}
    \text{corr}(\delta \xi(t), \delta \rho(t)) \simeq \frac{\xi_0^2 \rho_0^2}{L^4}\left[\frac{1}{2} \, (3 \cos^2\theta -1)\right],
\end{equation}
where the polar angle $\theta$ is the angle between the two detectors, since $\Vec{\xi}(t)$ is along the z-axis. Also, as previously mentioned, $\text{corr}( \xi(t),  \rho(t))=\text{corr}(\delta \xi(t), \delta \rho(t))$.

\begin{figure}
    \centering
    \includegraphics[width=0.45\textwidth]{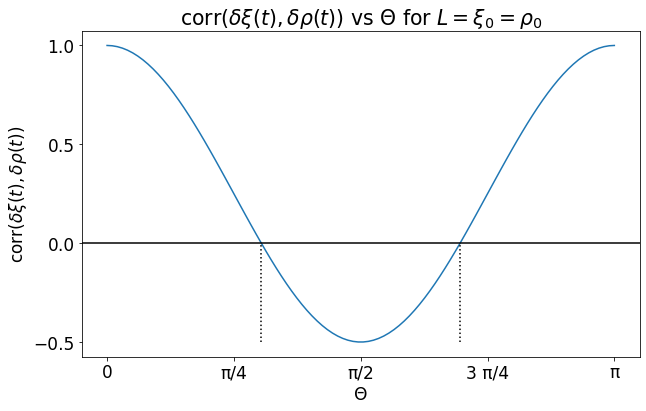}
    \caption{Plot of the correlation $\text{corr}(\delta \xi(t), \delta \rho(t))$ vs the polar angle $\theta$ for $L=\xi_0=\rho_0$ as described by Eq. ($\ref{main}$). When the detectors are aligned the noise correlation is $1$, while when they are orthogonal the noise correlation is $-1/2$. The correlation is zero at $\theta = \cos^{-1}\left(\pm1/\sqrt{3} \right)$, which correspond to roughly $54.7^{\circ}$ and $125.3^{\circ}$.}
    \label{corr vs theta}
\end{figure}

\subsection{Phenomenology}

Equation (\ref{main}) indicates that the correlation of the graviton noise in the two detectors is constant in time and it depends on the arm length of the detectors, the size of the system, and the angle between the orientation of detectors. Since the size of the system is always smaller than or equal to the arm length of the individual detectors, the factor that involves the $\xi$, $\rho$, and $L$ is never larger than one, and it is equal to one only if $L=\xi_0=\rho_0$. The closer the detectors are, the larger this factor is. On the contrary, this ratio becomes zero in the limit that the detectors are infinitely far away from each other. 

Figure \ref{corr vs theta} contains the plot of the correlation $\text{corr}(\delta \xi(t), \delta \rho(t))$ vs the polar angle $\theta$ for the special case $L=\xi_0=\rho_0$. As we can see, a maximum correlation of $1$ occurs when theta is $0$ or $\pi$, meaning that the correlation is maximized when the detectors are aligned. On the contrary, a correlation of $-1/2$ occurs when the detectors are orthogonal so that $\theta$ is $\pi/2$. The correlation is never below this value. Finally, the correlation is zero for values of $\theta$ of roughly $54.7^{\circ}$ and $125.3^{\circ}$.

\section{Using the Result}
\subsection{Designing a tabletop experiment}

We previously found that the variance of the noise in a detector of size $\xi$ is proportional to $\xi^2 \Lambda^4_\xi \sim \xi^{-2}$. This indicates that the noise is easier to observe if the arm length of the detector is small, since for a small detector the cutoff frequency is larger, and so more modes will contribute to the noise. Because of this, it seems advantageous to design a tabletop experiment in which small detectors are used, rather than large detectors like LIGO. For instance, using equations (\ref{amplitude}) and (\ref{var xi}), we can estimate that, if $\xi = 5$ m (which is roughly the size of the GQuEST experiment\cite{McCuller:2022hum, Vermeulen:2024vgl}) and $t = 10^3$ s, then the standard deviation of the noise in the usual Poincar\'e-invariant vacuum state is approximately $10^{-23} \; \text{m}$. This value would be smaller for larger values of $\xi$. Note that since the variance is proportional to $t^2$, it might be possible to detect the quantum-gravitational noise even without squeezing, for large values of $t$. We can also use equations (\ref{2-point function with integrals}) and (\ref{k}) to obtain an equation for the power spectrum of the quantum-gravitational noise
\begin{equation}
    S(\omega)= \frac{16 \hbar G \xi^2}{15 \pi^2 c^5} \omega,
\end{equation}
where $\xi$ is the arm length of the detector and we explicitly inserted the powers of $c$ that had been previously omitted. Evaluating this at $\xi = 5$ m and $\omega = \Lambda_\xi$, we obtain
\begin{equation}
    S(\Lambda_\xi)= \left(1.7*10^{-39} \; \text{m}/\sqrt{\text{Hz}} \right)^2
\end{equation}
Now, the power spectral densities of thermal noise and photon shot noise, which are the leading sources of noise in GQuEST, are estimated to be \cite{Vermeulen:2024vgl}
\begin{equation}
\begin{split}
    S_\text{thermal}(\omega \sim \Lambda_\xi) &= \left(10^{-21} \; \text{m}/\sqrt{\text{Hz}} \right)^2 \\
    S_\text{shot}(\omega \sim \Lambda_\xi) &= \left(6*10^{-19} \; \text{m}/\sqrt{\text{Hz}} \right)^2
\end{split}
\end{equation}
Although these estimates indicate that the quantum-gravitational noise is much smaller than the noise from other sources, this does imply that it cannot be detected. In fact, as discussed above, the variance of the quantum-gravitational noise grows with time, while this should not be the case for the other sources of noise. Additionally, the quantum-gravitational noise is correlated in two different detectors, while the thermal and photon shot noise are not. Because of this, it should be possible to detect the quantum-gravitational noise even if it is not as large as the uncorrelated noise due to other sources.  

\begin{figure}
    \centering
    \subfloat[Case in which the noise is correlated in all detectors. In this case, the correlated noise is seismic noise.]{
        \includegraphics[width=0.45\textwidth]{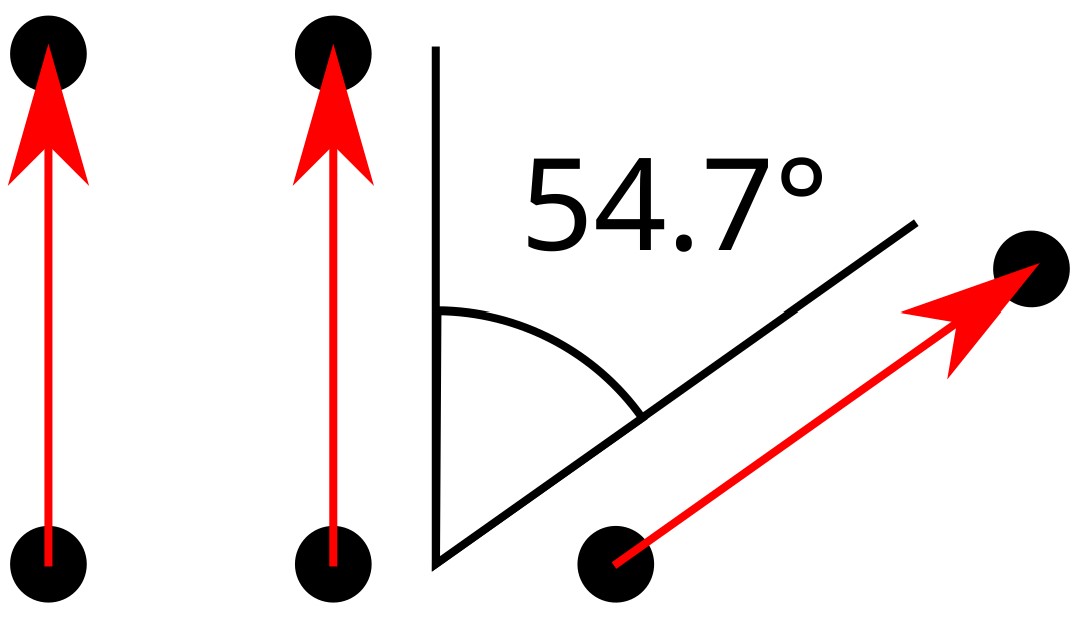}
        \label{correlated}
    }
    \hfill
    \subfloat[Case in which the noise is correlated only in the two aligned detectors. In this case, the correlated noise in the aligned detectors is due to quantum gravity. ]{
        \includegraphics[width=0.45\textwidth]{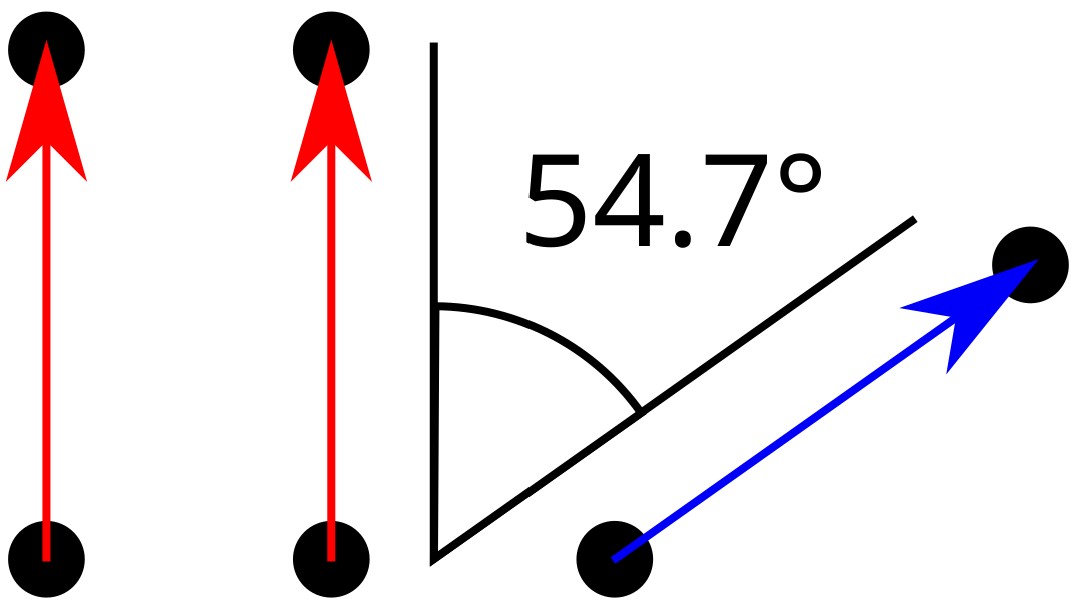}
        \label{not_correlated}
    }
    \caption{Proposed apparatus to detect the quantum-gravitational noise. Two detectors should be aligned with each other and a third detector should make an angle of either $54.7^{\circ}$ or $125.3^{\circ}$ with the other detectors. All three detectors should be as close to each other as possible. In this configuration, the quantum-gravitational noise in the parallel detectors should have a correlation approximately equal to 1, while the quantum-gravitational noise in the nonparallel detector has a correlation equal to zero. If the noise in all three detectors is correlated, as shown in the image on the left, then the noise is not due to quantum gravity. If the noise in the parallel detectors is correlated and the noise in the nonparallel detector is not correlated with the noise in the other detectors, as shown in the image on the right, then the noise detected in the parallel detectors is indeed due to quantum gravity.}
    \label{tabletop_experiment}
\end{figure}

Equation (\ref{main}), our main result, indicates that the correlation of the quantum gravity noise in two different detectors is maximized when the detectors are aligned and close to each other. Thus, a possible experiment might involve two aligned detectors with arm length on the order of a meter placed a few centimeters away from each other. However, if the detectors are this close to each other, there could be correlated background noise, such as seismic noise. Now, the correlated component of the background noise presumably does not have the same angular dependence of the quantum-gravitational noise. Thus, to distinguish quantum-gravitational noise from correlated background noise, we propose adding a third detector arm placed at a $54.7^{\circ}$ or $125.3^{\circ}$ angle with the other two parallel detectors, as shown in Fig. \ref{tabletop_experiment}. The advantage of this configuration is that the quantum-gravitational noise in the third detector should have a correlation of zero with the quantum-gravitational noise in the parallel detectors. With this setup, it would be possible to distinguish the quantum-gravitational noise both from correlated background noise (such as seismic noise) as well as from uncorrelated background noise (such as thermal, electronic, or shot noise).

For example, suppose that we collect some data and observe that the noise in all three detectors is correlated, as in Fig. \ref{correlated}. Then, we would conclude that the observed noise is not due to quantum gravity, but rather is some other correlated noise, such as seismic noise. By contrast, if we observe that only the noise in the parallel detectors is correlated, as in Fig. \ref{not_correlated}, then we would conclude that the correlated noise in the parallel detectors is indeed due to quantum gravity rather than to correlated or uncorrelated background noise. Of course, we cannot simply use two nonparallel detectors because then we might not be able to confirm whether the uncorrelated noise is actually due to quantum gravity, rather than to other sources of uncorrelated noise. 

\subsection{Detecting quantum gravity noise with an interferometer}

So far we have modeled our detectors as two mirrors, where the arm length is given by the mirrors' separation. However, real gravitational wave detectors are much more complex. We would now like to show how to apply our result to a gravitational wave interferometer. 

\begin{figure}
    \centering
    \includegraphics[width=0.45\textwidth]{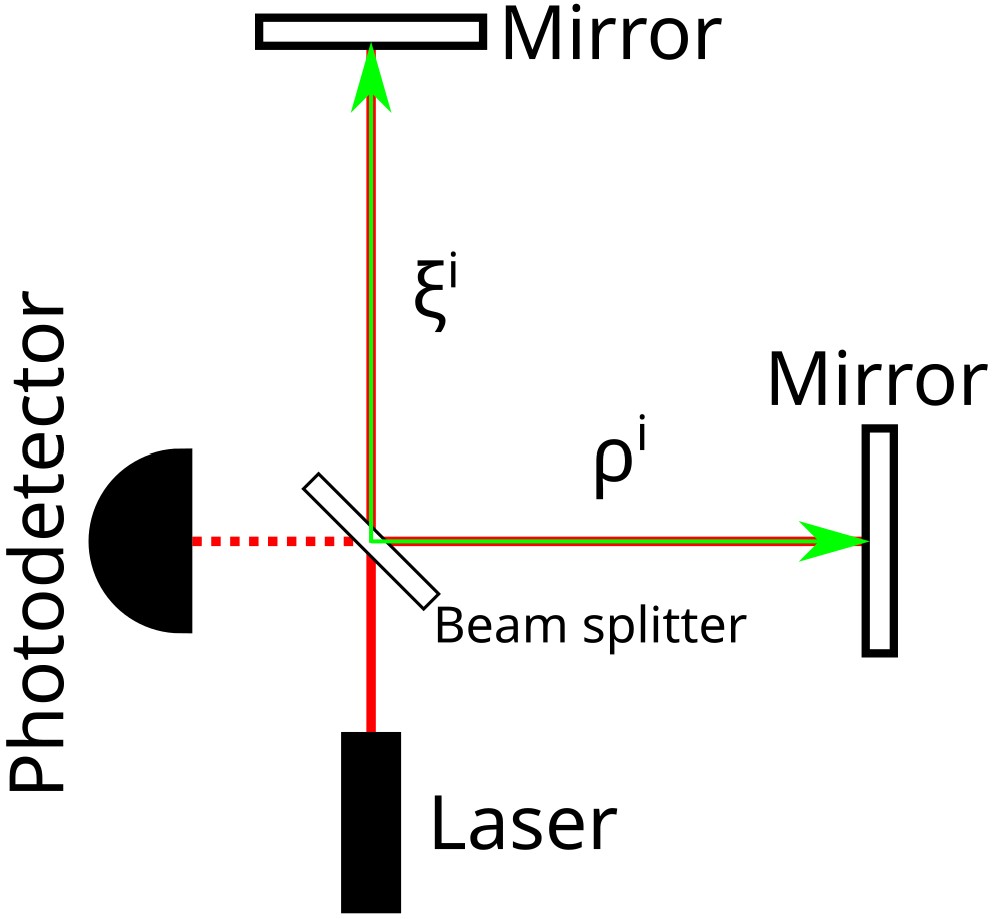}
    \caption{Basic layout of a gravitational wave interferometer. The beam splitter and the two mirrors can be modeled as two simple ``detectors," where the beam splitter is the reference mass and the arms are $\xi^i$ and $\rho^i$.}
    \label{interferometer}
\end{figure}

A basic interferometer is constituted by a laser beam, two mirrors, a beam splitter, and a photodetector, as shown in Fig. \ref{interferometer}. When it reaches the beam splitter, the laser is separated into two beams in phase with each other. The beams travel down the arms and back to the beam splitter, where they interfere with each other. If fluctuations changed the relative length of the two arms, the beams will not be in phase anymore once they reach the beam splitter. If this phase difference is large enough, the photodetector will be able to measure it. Although the laser also travels from its source to the beam splitter, and from the beam splitter to the photodetector, we are not interested in the fluctuations that occur along these paths because they do not contribute to the relative phase shift of the split beams in the arms. On the contrary, we are interested in how the quantum-gravitational noise affects the position of the mirrors relative to the beam splitter, since the fluctuations in the position of the mirrors are responsible for the relative phase shift measured by the photodetector. To quantify this effect, we can model the beam splitter and the mirrors as three stationary masses, take the beam splitter as the reference, and consider the arms of the interferometer as two simple ``detectors" of the same kind that we analyzed in the previous sections. Let the arms of the ``detectors" be $\xi^i(t)$ and $\rho^i(t)$. In a typical interferometer, the arms have the same unperturbed length and are perpendicular to each other. In Eq. (\ref{main}), this corresponds to $\xi_0 = \rho_0$, $L = \sqrt{(\xi_0)^2+(\rho_0)^2}=\sqrt{2}\,\xi_0$, and $\theta = \pi/2$. Using these values, we find that the correlation of the noise in the arms of the interferometer is 
\begin{equation}
\begin{split}
    \text{corr}(\delta \xi(t), \delta \rho(t)) \simeq& \frac{\xi_0^2 \xi_0^2}{(\sqrt{2}\,\xi_0)^4}\left[\frac{1}{2} \, \left(3 \cos^2\left(\frac{\pi}{2}\right) -1\right)\right] \\=& \frac{1}{4} \left[ - \frac{1}{2} \right] = - \frac{1}{8}
\end{split}
\end{equation}
Thus, the correlation is fairly small but it is negative. The fact that the correlation is negative is beneficial to the detection of the quantum-gravitational noise because it means that, on average, if one mirror moves away from the beam splitter, the other mirror moves closer to the beam splitter. This will produce a difference in the length of the arms, which might lead to a measurable phase difference at the beam splitter. On the contrary, if the correlation was exactly $1$, we would not be able to detect the noise because, even if the fluctuations were large, the change in the length of the arms would be identical, and so the beams would be in phase with each other once they travel back to the beam splitter. 

We conclude this section by pointing out that the previously described setup (arms with equal length and perpendicular to each other) is the ideal configuration to detect the quantum-gravitational noise using a single gravitational wave interferometer. In fact, the closer the noise correlation is to $-1$, the larger the signal observed by the photodetector. As shown in Fig. \ref{corr vs theta}, we know that the angle that makes the correlation closer to $-1$ is $90^\circ$. Additionally, assuming that the detector arms cannot overlap, the factor $\xi_0^2 \rho_0^2 / L^4$ is maximized for $\xi_0 = \rho_0$. Putting everything together, this gives the closest possible value to $-1$, which we previously determined to be $-1/8$.

\section{Discussion}
We have calculated the correlation in the fundamental quantum-gravitational noise between a pair of gravitational wave detector arms. While the amplitude of the noise depends on the quantum state of the gravitational field, the correlation has a characteristic form, at least for states with rotational symmetry such as the vacuum state. We found, unsurprisingly, that the correlation falls off as the arms are separated in space. The correlation otherwise depends only on the angle between the orientations of the detector arms: it is maximal when they are aligned and zero for $\cos^2 \theta = 1/3$ i.e. for an angle of $54.7^{\circ}$ or $125.3^{\circ}$.
This suggests that, in order to maximally distinguish the gravitational noise from other sources of noise, an ideal configuration of gravitational wave detectors could consist of three arms, two of which are aligned with the third at an angle of $54.7^{\circ}$ or $125.3^{\circ}$. The detectors should be close enough to maximize the correlation while still being far enough apart to eliminate correlated noise from other environmental sources. Under these conditions, noise that was correlated between the two aligned detectors but uncorrelated with the third one would likely be quantum-gravitational noise. Conversely, our result can be used to identify the correlated environmental noise by assessing whether there is any correlation in the noise of the detectors placed at an angle of $54.7^{\circ}$ or $125.3^{\circ}$: if such correlations are observed, then the observed noise cannot be due to quantum gravity. Finally, we explained how our result can be applied to a realistic gravitational wave interferometer. Since the quantum-gravitational noise in the arms of the detector has a correlation of $-1/8$, it is possible to detect it using an interferometer, provided that its variance is large enough.\\

\noindent
{\bf ACKNOWLEDGMENTS}\\
We thank Sang-Eon Bak for helpful discussions. M.P. is supported in part by Heising-Simons Foundation Grant No. 2021-2818, Department of Energy Grant No. DE-SC0019470 and Government of India DST VAJRA Faculty Scheme VJR/2017/000117.

\appendix
\section{APPENDIX}
Here are the calculations to derive the variance of $\delta \rho(t)$ given in equation ($\ref{var rho}$).

Since $\expval{\delta \rho(t)}_s=0$, we have 
\begin{align}
\begin{split}
    &\text{Var}(\delta \rho(t)) =  \expval{\delta \rho(t)\delta \rho(t)}_s \\
    =&\expval{\left[ \delta \Vec{\rho}(t) \cdot \Vec{\rho_0}/\rho_0\right]\left[ \delta \Vec{\rho}(t) \cdot \Vec{\rho_0}/\rho_0\right]}_s \\
    =& \Bigl[ \expval{\delta \rho^1(t)\delta \rho^1(t)}_s \left(\rho^1_0\right)^2 + \expval{\delta \rho^2(t)\delta \rho^2(t)}_s \left(\rho^2_0\right)^2 \\
    +&\expval{\delta \rho^3(t)\delta \rho^3(t)}_s \left(\rho^3_0\right)^2  +2\expval{\delta \rho^1(t)\delta \rho^2(t)}_s \rho^1_0\rho^2_0 \\
    +&2\expval{\delta \rho^1(t)\delta \rho^3(t)}_s \rho^1_0\rho^3_0+2\expval{\delta \rho^2(t)\delta \rho^3(t)}_s \rho^2_0\rho^3_0\Bigr]/\left(\rho_0\right)^2
\end{split}
\end{align}
Thus, we need to evaluate the 2-point functions involving all possible combinations of the components of $\delta \rho^i(t)$. For example, 
\begin{align}
\begin{split}
    &\expval{\delta \rho^1(t) \delta \rho^1(t)}_s = \expval{\left[\delta \zeta^1(t)-\delta \eta^1(t) \right]\left[\delta \zeta^1(t)-\delta \eta^1(t) \right]}_s \\
    &= \expval{\delta \zeta^1(t)\delta \zeta^1(t)}_s + \expval{\delta \eta^1(t)\delta \eta^1(t)}_s - 2 \expval{\delta \zeta^1(t)\delta \eta^1(t)}_s 
\end{split}
\end{align}
or
\begin{align}
\begin{split}
    \expval{\delta \rho^1(t) \delta \rho^2(t)}_s =& \expval{\left[\delta \zeta^1(t)-\delta \eta^1(t) \right]\left[\delta \zeta^2(t)-\delta \eta^2(t) \right]}_s \\
    =& \expval{\delta \zeta^1(t)\delta \zeta^2(t)}_s + \expval{\delta \eta^1(t)\delta \eta^2(t)}_s \\ 
    & -\expval{\delta \zeta^1(t)\delta \eta^2(t)}_s -\expval{\delta \zeta^2(t)\delta \eta^1(t)}_s 
\end{split}
\end{align}
From these two examples, we infer that in order to evaluate $\expval{\delta \rho^i(t)\delta \rho^j(t)}_s$, we need to evaluate $\expval{\delta \zeta^i(t) \delta \eta^j(t)}_s$ for all possible values of $i$ and $j$. 

Following the same procedure shown in the previous sections, it is easy to obtain equations for $\delta \eta^i(t)$ and $\delta \zeta^i(t)$ analogous to ($\ref{xi_1}$). We can then combine these and obtain
\begin{equation}
\begin{split}
    \langle\delta \zeta^i(t) & \delta \eta^j(t')\rangle_s = (2  \alpha)^2 \, \delta^{ik} \delta^{jl} \zeta_0^m \eta_0^n \\& \int^t_0 dt' \int^{t'}_0 dt'' \int^{t^{(3)}}_0 dt^{(4)} \int^{t^{(4)}}_0 dt^{(5)} K^\Psi_{kmln}(t,t'),
\end{split}
\end{equation}
where the noise kernel is the same as the one from Eq. ($\ref{k}$). Then, it follows that 
\begin{equation}
\begin{split}
    \langle \delta \zeta^i(t)  \delta \eta^j(t)\rangle_s =& A(\Lambda,t) \delta^{ik} \delta^{jl} \zeta_0^m \eta_0^n \\&[-2 \delta_{km} \delta_{ln} +3 (\delta_{kl} \delta_{mn} +\delta_{kn}\delta_{ml})],
\end{split}
\end{equation}
where $A(\Lambda,t)$ is given in Eq. ($\ref{amplitude}$). We can now evaluate the 2-point functions for all values of $i$ and $j$.

Suppose that $i=1$. Then, 
\begin{equation}
\begin{split}
    \expval{\delta \zeta^1(t) \delta \eta^j(t)}_s =& A(\Lambda,t) \delta^{11} \delta^{jl} \zeta_0^m \eta_0^n \\&[-2 \delta_{1m} \delta_{ln} +3 (\delta_{1l} \delta_{mn} +\delta_{1n}\delta_{ml})]
\end{split}
\end{equation}
Using this, we obtain 
\begin{align}
\begin{split}
    j=1 : &\expval{\delta \zeta^1(t) \delta \eta^1(t)}_s /A(\Lambda,t) \\&= \,  \delta^{11} \delta^{11} \zeta_0^m \eta_0^n [-2 \delta_{1m} \delta_{1n} +3 (\delta_{11} \delta_{mn} +\delta_{1n}\delta_{m1})] \\
    &= -2 \zeta^1_0 \eta^1_0 + 3 \left(\Vec{\zeta}_0 \cdot \Vec{\eta}_0 + \zeta^1_0 \eta^1_0 \right) \\
    &= \, 3 \Vec{\zeta}_0 \cdot \Vec{\eta}_0 +\zeta^1_0 \eta^1_0 \\
    j=2 : &\expval{\delta \zeta^1(t) \delta \eta^2(t)}_s /A(\Lambda,t) \\&= \,  \delta^{11} \delta^{22} \zeta_0^m \eta_0^n [-2 \delta_{1m} \delta_{2n} +3 (\delta_{12} \delta_{mn} +\delta_{1n}\delta_{m2})] \\
    &= -2 \zeta_0^1 \eta_0^2 +3 \zeta_0^2 \eta_0^1 \\
    j=3 : &\expval{\delta \zeta^1(t) \delta \eta^3(t)}_s /A(\Lambda,t) \\&= \,  \delta^{11} \delta^{33} \zeta_0^m \eta_0^n [-2 \delta_{1m} \delta_{3n} +3 (\delta_{13} \delta_{mn} +\delta_{1n}\delta_{m3})] \\
    &= -2 \zeta_0^1 \eta_0^3 +3 \zeta_0^3 \eta_0^1 
\end{split}
\end{align}
From this example, we infer that 
\begin{equation}
    \expval{\delta \zeta^i(t) \delta \eta^j(t)}_s /A(\Lambda,t) = \begin{cases} \, 3 \Vec{\zeta}_0 \cdot \Vec{\eta}_0 +\zeta^i_0 \eta^j_0  &\text{if} \: \; i=j \\
    -2 \zeta_0^i \eta_0^j +3 \zeta_0^j \eta_0^i  &\text{if} \: \; i\neq j
    \end{cases}
\end{equation}
Also, completely analogous rules apply to $\expval{\delta \zeta^i(t) \delta \zeta^j(t)}_s$ and $\expval{\delta \eta^i(t) \delta \eta^j(t)}_s$.

Using these we can now evaluate $\expval{\delta \rho^i(t)\delta \rho^j(t)}_s$ for all values of $i$ and $j$. For example,
\begin{align}
\begin{split}
    &\langle\delta \rho^1(t)\delta \rho^1(t)\rangle_s /A(\Lambda,t) \\&= \bigl[\expval{\delta \zeta^1(t)\delta \zeta^1(t)}_s + \expval{\delta \eta^1(t)\delta \eta^1(t)}_s \\&\quad - 2 \expval{\delta \zeta^1(t)\delta \eta^1(t)}_s\bigr]/A(\Lambda,t) \\
    &= \, 3 \left( \zeta_0 \right)^2 + \left( \zeta_0^1 \right)^2 + 3 \left( \eta_0 \right)^2 + \left( \eta_0^1 \right)^2 -6 \Vec{\zeta}_0 \cdot \Vec{\eta}_0 - 2 \zeta_0^1 \eta_0^1 \\
    &= \, 3 \left( \Vec{\zeta}_0 - \Vec{\eta}_0 \right)^2 + \left(\zeta_0^1 - \eta_0^1\right)^2 \\
    &= \, 3 \left( \rho_0 \right)^2 + \left( \rho_0^1 \right)^2
\end{split}
\end{align}
and 
\begin{align}
\begin{split}
    &\expval{\delta \rho^1(t)\delta \rho^2(t)}_s /A(\Lambda,t) \\&= \bigl[\expval{\delta \zeta^1(t)\delta \zeta^2(t)}_s + \expval{\delta \eta^1(t)\delta \eta^2(t)}_s \\ 
    &\quad-\expval{\delta \zeta^1(t)\delta \eta^2(t)}_s -\expval{\delta \zeta^2(t)\delta \eta^1(t)}_s \bigr]/A(\Lambda,t) \\
    &= \, \left(-2 \zeta_0^1 \zeta_0^2 + 3 \zeta_0^2 \zeta_0^1 \right) + \left(-2 \eta_0^1 \eta_0^2 + 3 \eta_0^2 \eta_0^1 
    \right) \\
    & \quad- \left(-2 \zeta_0^1 \eta_0^2 + 3 \zeta_0^2 \eta_0^1 \right) - \left(-2 \zeta_0^2 \eta_0^1 + 3 \zeta_0^1 \eta_0^2 \right) \\
    &= \, \zeta_0^1 \zeta_0^2 + \eta_0^1 \eta_0^2 - \zeta_0^1 \eta_0^2 - \zeta_0^2 \eta_0^1 \\
    &= \left( \zeta_0^1 - \eta_0^1 \right) \left( \zeta_0^2 - \eta_0^2 \right) \\
    &= \, \rho_0^1 \rho_0^2 
\end{split}
\end{align}
From these calculations, we infer the following rules
\begin{equation}
    \expval{\delta \rho^i(t) \delta \rho^j(t)}_s /A(\Lambda,t) = \begin{cases} \, 3 \left( \rho_0 \right)^2 + \left( \rho_0^i \right)^2  &\text{if} \: \; i=j \\
    \, \rho_0^i \rho_0^j  &\text{if} \: \; i\neq j
    \end{cases}
\end{equation}

Finally, we can use these rules to finish evaluating $\text{Var}(\delta \rho(t))$. We obtain
\begin{align}
\begin{split}
    &\text{Var}(\delta \rho(t)) \\&= \Bigl[ \expval{\delta \rho^1(t)\delta \rho^1(t)}_s \left(\rho^1_0\right)^2 + \expval{\delta \rho^2(t)\delta \rho^2(t)}_s \left(\rho^2_0\right)^2 \\
    &+\expval{\delta \rho^3(t)\delta \rho^3(t)}_s \left(\rho^3_0\right)^2  +2\expval{\delta \rho^1(t)\delta \rho^2(t)}_s \rho^1_0\rho^2_0 \\
    &+2\expval{\delta \rho^1(t)\delta \rho^3(t)}_s \rho^1_0\rho^3_0+2\expval{\delta \rho^2(t)\delta \rho^3(t)}_s \rho^2_0\rho^3_0\Bigr]/\left(\rho_0\right)^2 \\
    &= \Bigl[ \left( 3 \left( \rho_0 \right)^2 + \left( \rho_0^1 \right)^2 \right) \left(\rho^1_0\right)^2 + \left( 3 \left( \rho_0 \right)^2 + \left( \rho_0^2 \right)^2 \right) \left(\rho^2_0\right)^2 \\
    &+\left( 3 \left( \rho_0 \right)^2 + \left( \rho_0^3 \right)^2 \right) \left(\rho^3_0\right)^2  +2 \, \rho_0^1 \rho_0^2 \, \rho^1_0\rho^2_0 \\
    &+2\, \rho_0^1 \rho_0^3 \, \rho^1_0\rho^3_0+2 \, \rho_0^2 \rho_0^3 \, \rho^2_0\rho^3_0\Bigr]A(\Lambda,t)/\left(\rho_0\right)^2 \\
    &= \biggl[ 3 \left(\rho_0\right)^2\left[\left(\rho_0^1\right)^2+\left(\rho_0^2\right)^2+\left(\rho_0^3\right)^2  \right]\\&+\left[\left(\rho_0^1\right)^2+\left(\rho_0^2\right)^2+\left(\rho_0^3\right)^2 \right]^2 \biggr]A(\Lambda,t)/\left(\rho_0\right)^2 \\
    &= \, 4 \left(\rho_0\right)^2 A(\Lambda,t).
\end{split}
\end{align}

%%%%%%%%%%%%%%%%%%%%%%%%%%%%%%%%%%%%%%
%    END
%%%%%%%%%%%%%%%%%%%%%%%%%%%%%%%%%%%%%%

%\bibliography{bibtex}
%\bibliographystyle{JHEP}

%apsrev4-2.bst 2019-01-14 (MD) hand-edited version of apsrev4-1.bst
%Control: key (0)
%Control: author (8) initials jnrlst
%Control: editor formatted (1) identically to author
%Control: production of article title (0) allowed
%Control: page (0) single
%Control: year (1) truncated
%Control: production of eprint (0) enabled
%

\end{document}